# Optimization of Vehicle Trajectories Considering Uncertainty in Actuated Traffic Signal Timings

Amr Shafik, Seifeldeen Eteifa, and Hesham A. Rakha, *Fellow, IEEE*

*Abstract*— This paper introduces a robust optimal green light speed advisory system for fixed and actuated traffic signals when a probability distribution is provided. These distributions represent the domain of possible switching times from the Signal Phasing and Timing (SPaT) messages. The system finds the least-cost vehicle trajectory using a computationally efficient A-star algorithm incorporated in a dynamic programming procedure to minimize the vehicle's total fuel consumption. Constraints are introduced to ensure that vehicles do not, collide with other vehicles, run red lights, or exceed a maximum vehicular jerk for passenger comfort. Results of simulation scenarios are evaluated against comparable trajectories of uninformed drivers to compute fuel consumption savings. The proposed approach produced significant fuel savings compared to the uninformed driver amounting to 37 percent on average for deterministic SPaT and 28 percent for stochastic SPaT. A sensitivity analysis is performed to understand how the degree of uncertainty in SPaT predictions affects the optimal trajectory's fuel consumption. The results present the required levels of confidence in these predictions to achieve most of the possible savings in fuel consumption. Specifically, the proposed system can be within 85% of the maximum savings if the timing error is (±3.3 seconds) at a 95% confidence level. They also emphasize the importance of more reliable SPaT predictions the closer the time to green is relative to the time the vehicle is expected to reach the intersection given its current speed.

## I. Introduction

THE global energy system is severely affected by the crisis caused by the shortage of fuel supply. Due to the excessive use of fossil fuels, climate change exacerbated resulting in more heat waves that impose global health threats [1]. The transportation sector consumes about 28% of the total energy consumed by the US (based on 2021 statistics), of which 90% comes from petroleum products [2]. It also contributes 27% of the total greenhouse gas emissions resulting in air quality degradation and climate change [3-5]. As a result, there is a global trend to reduce the consumption of fossil fuels in the transportation sector by raising awareness of the crisis, using energy-efficient systems, and using alternative energy sources.

Eco-driving is an efficient and cost-effective system that can significantly reduce fuel consumption. Eco-driving has been proven to achieve fuel savings of up to 45% [6]. Eco-driving at signalized intersections is known as Green Light Optimal Speed Advisory (GLOSA). Traffic signals impose travel time delays due to the vehicles stopping at red lights. Traffic signals also impose additional fuel consumption due to idling and aggressive acceleration. This is estimated at 2.8 million gallons in the U.S. [7, 8].

Research was carried out to improve signal control and eco-driving systems to reduce fuel losses at intersections using Infrastructure to Vehicle communication (I2V) technologies [9]. Q-learning, artificial neural networks, and fuzzy logic were all employed to optimize traffic signal controllers [10]. Dynamic eco-driving models developed vastly in the last decade [11]. These models addressed fuel consumption due to unnecessary acceleration upstream of signalized intersections. This allows vehicles to reduce idling at red lights using signal phasing and timing (SPaT) information.

In fixed-time signal controllers, SPaT messages indicate exact information about signal switching times allowing vehicles to deterministically optimize their speed profile. This optimization significantly reduces fuel consumption levels. Kamalanathsharma and Rakha [12] developed an Eco-cooperative adaptive cruise control (Eco-CACC) system that uses I2V communication to recommend fuel-efficient speeds for the vehicle approaching a fixed-time signal controller. The algorithm explicitly minimized the fuel consumption as an objective function. Using a modified A* algorithm, the problem was solved through dynamic programming to achieve average fuel savings of up to 30%.

In the case of actuated traffic signal controllers, which are widely deployed in the U.S., a reliable switching time is unavailable in SPaT messages due to real-time signal control updates. Previous research efforts utilized the available data in SPaT messages to predict a most likely switching time or probabilistic distributions for switching times. These studies used traffic signal parameters, time of day, detector data, vehicle and pedestrians actuation data, and controller logic to predict signal switching times [13].

*Eco-driving at Actuated Traffic Signals*

Given uncertain switching time predictions, significant levels of fuel savings can still be achieved as indicated by the work of Mahler and Vahidi (14). They developed an algorithm to plan the vehicle's optimal velocity while approaching actuated traffic signals (14). SPaT predictions used in the algorithm are based on historical data and real-time phase information. This study used a simplified cost function representing fuel consumption levels rather than using an actual fuel consumption model to avoid the associated computational complexity. Results showed that fuel economy can be achieved up to 6% over the uninformed driver for actuated signals.

Another stochastic eco-driving control system was developed by Sun et al. (15). The system deals with a vehicle proceeding through multiple actuated traffic signal controllers with uncertain SPaT information. The problem was formulated as a chance-constrained stochastic program, in which an effective red-light duration (ERD) concept was introduced to capture the randomness in the traffic signal switching times. Using Dynamic Programming (DP), about 50-57% savings in fuel consumption were achieved with minimum sacrifice in arrival times. The major drawback of this study is using general probability density functions to account for uncertain switching times for different traffic conditions rather than utilizing the real-world SPaT information. Additionally, the used fuel consumption model has a high level of computational complexity due to using engine and transmission torque and the gear number in the calculations, making it inefficient for real-time application. Overall, previous literature indicates it is feasible to enhance eco-driving at signalized intersections using SPaT with no major changes to the current infrastructure.

## II. PROBLEM STATEMENT

Various researchers devised methods for dealing with the uncertainty in the SPAT predictions and making eco-driving control decisions that achieve fuel savings despite the uncertainty in the information. However, the literature lacks information on the extent of the effect of uncertainty in SPAT information on the fuel savings achievable. This makes it difficult to assess when a SPAT prediction is good enough or when the error in SPAT prediction is within reasonable limits. Knowledge of the effect of the error in the SPAT information on the achievable fuel consumption can help assess the SPAT most likely time prediction validity from an application standpoint. This can also help with the choice of statistical or machine learning model choices. Better models can be chosen by looking at their prediction error distributions and having a reference on how this error is expected to affect the fuel consumption savings given optimal control during eco-driving. This study investigates the effect of the degree of uncertainty in SPAT predictions on the achievable fuel savings given optimal vehicle control (Fig. 1).

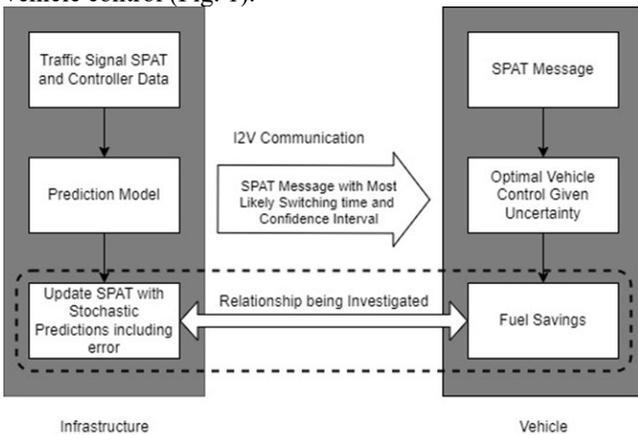

**Fig. 1.** SPAT I2V Communication for Actuated Signals and Study Motivation

*Study Contribution*

This study has several contributions to the existing body of knowledge and potential benefits to practitioners including infrastructure operators and automotive original equipment manufacturers (OEMs). The drawbacks found in the literature mentioned above are also addressed in this study. The main contributions can be summarized as follows:
- Extending the deterministic vehicle control algorithm proposed by Kamalanathsharma and Rakha (*14*) to a stochastic algorithm, which is explicitly minimizing fuel consumption as an objective function. In addition, the system also accounts for vehicle jerk constraints for passenger comfort.
- Comparing the fuel consumption savings resulted from the proposed system with actual field trajectories for the case of uninformed drivers.
- Detailing a framework for the safe application of GLOSA given a stochastic prediction.
- Identifying benchmarks for practitioners attempting to use statistical or machine learning models to predict SPaT most likely switching times regarding both model bias and variance required to achieve the benefits of GLOSA.
- Providing OEMs with an objective quantification of the effect of level of confidence in SPaT predictions that they can use to reliably control their vehicles.

## III. METHODOLOGY

In this paper, we extend the work of Kamalanathsharma and Rakha (14) Kamalanathsharma and Rakha (12) deterministic vehicle control algorithm by constraining the vehicle acceleration level with comfortable levels of jerk, as well as by including the case of actuated signal controllers. The problem of optimizing vehicle trajectory is formulated as a robust optimal control problem, in which the real-time uncertain signal switching time is used to calculate and update the optimal vehicle control policy in real-time. The updated policy is based on new information received through I2V communication. A switching time probabilistic distribution is transmitted to the vehicle in real-time. The optimization system explicitly minimizes vehicle fuel consumption as an objective function. Dynamic Programming (DP) and A* algorithm are used to solve the problem numerically tackling computational complexity. The algorithm's robustness towards uncertain signal information is ensured through a risk assessment procedure to prevent red light violations. The eco-driving approach simulation results are compared to the case of an uninformed driver approaching a traffic signal without prior information about the switching time as a baseline. The uninformed driver data was retrieved from a field experiment conducted at the Virginia smart road test facility at the Virginia Tech Transportation Institute (VTTI) (16).

This study adopts a four-step research methodology described as follows:
1. Define the eco-driving problem at hand with all different analysis scenarios.

2. Develop a holistic eco-driving system that takes stochasticity and changing information into account without violating any of the constraints imposed by the vehicle and the traffic signal.
3. Evaluate the performance of the system under:
   a. The different pre-defined scenarios
   b. Various levels of bias and variance are introduced in the predictions.
4. Identify the effect of the vehicle's initial speed on the overall system performance.

*A. Base Analysis Scenarios*

The base scenario is defined as follows: we have a vehicle traveling on a single-lane approach, with no other traffic, with a free-flow speed ($v_0$) of 40 mph (64.4 kph) which is common for urban arterials. As shown in Fig. 2, $X_0$ is the beginning position, which is 250 meters upstream of the stop bar of an actuated traffic signal controller located at the position $X_M$. The choice of 250 meters is due to the reliable range of Dynamic Short-Range Communication (DSRC) and C-V2X existing technologies beyond which significant message losses and interference occur [14, 15]. The vehicle should accelerate to the desired speed at or before position $X_N$, which is 180 meters downstream of the stop bar. The vertical grade (G) is chosen to be 3% either uphill or downhill. This is the same setup used in the field experiment [16], which allows us to use field data collected about uninformed drivers as a baseline.

The assumption is that the vehicle is within the communication range of an actuated traffic signal controller. The signal is assumed to be transmitting SPaT information to the vehicle in real-time. Given the stochastic nature of actuated signals, SPaT information will contain uncertain most likely signal switching times from red to green in the form of a probability distribution of the switching time or a confidence level in the most likely switching time. Assuming a normal distribution for the errors, a confidence level can be used to infer a standard deviation and a mean for the distribution.

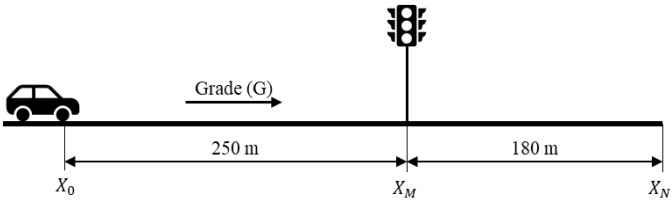

**Fig. 2.** Problem Setup

*B. Additional Scenarios*

The same problem setup is applied across multiple different scenarios. As shown in TABLE I, we have the uninformed driver (Scenario I), in which the vehicle driver approaches a traffic signal that is currently red. The driver is not aware of the traffic signal switching time, which takes one value from the set of switching times (10, 15, 20, or 25 seconds). In this scenario, the vehicle trajectory data is retrieved from the previous field experiment conducted at the Smart road.

In simulated scenarios II and III, the vehicle is provided with signal switching time information to optimize the vehicle trajectory. In scenario II, a fixed time signal is simulated by providing the vehicle with the exact switching time. A deterministic vehicle trajectory optimization algorithm is used to plan the acceleration/deceleration policy. In scenario III, an actuated traffic signal controller is simulated by providing the vehicle with stochastic switching times. The stochastic switching times are sampled from normal distributions at each analysis time step. The distribution mean is the true switching time plus an error bias value, and the standard deviation (STD) represents the variability in switching time provided by prediction algorithms. Different values of the bias and standard deviation are used to account for the switching time prediction quality levels. This uncertainty is inherent in the stochastic nature of the switching time prediction in actuated traffic signals.

TABLE I
ANALYSIS SCENARIOS

| Scenario | Description | Directions | Initial Speed (mph) | Switching Times (TTG) (sec) |
|---|---|---|---|---|
| I (Baseline) | Uninformed Driver | 1 (Downhill) or 2 (Uphill) | 40 | 10, 15, 20, 25 |
| II | Deterministic SPaT Information | | | |
| III | Stochastic Realtime SPaT Information | | | |

IV. STOCHASTIC PROBLEM STATEMENT & FORMULATION

To develop an optimal trajectory planning tool for green light optimal speed advisory (GLOSA), the problem is formulated as a stochastic optimal control problem. The exact signal switching time is assumed to be unknown to the system. Instead, the most likely or a probabilistic distribution of switching times is used to plan the trajectory. We assume that the vehicle receives uncertain signal information in real-time as it approaches the intersection at each period $\Delta t$.

The problem is identified as follows; let $x(t)$ be a discretized distance variable that belongs to the space $\mathbb{X} \in \mathbb{R}^n$. $v(t, x(t))$ is the speed policy variable that belongs to the policy space $\mathbb{V} \in \mathbb{R}^m$, which is dependent on either the throttle or the deceleration level as a decision variable. $\mathcal{V}_t$ is the admissible speed policy space, which is constrained by the vehicle dynamics, acceleration, jerk, and other system control constraints, such that the pair $(x, v) \in \mathbb{K}(t)$ for all $t = t_0, \ldots, t_f$, where $\mathbb{K}$ is the time-changing constraints set $\mathbb{K}(t) \subset \mathbb{X} \times \mathbb{V}$. The objective function of the problem is to minimize the total fuel consumption given the speed control policy $FC_{t,x}(v)$ (1), where $FC_{t,x}$ is the instantaneous fuel consumption at time $t$ at position $x$.

$$\underset{v \in \mathcal{V}_t}{\text{Min}} FC_{t,x}(v) \quad \forall\, t, x \in \mathbb{K}(t) \qquad (1)$$

The constraints included in the set $\mathbb{K}$ are illustrated as follows: Equation (2) shows the upstream distance constraint, where the vehicle traveled distance $X_s$ is less than or equal to the distance from the initial position to the intersection stop bar $X_M$. The upstream distance $X_s$ represents the distance traveled from the vehicle's initial position $t_0$ until the time when the signal switches to green, and in this case, the expected value of the switching time $t_s$. Note that using the expected value denotes a risk-neutral attitude, where the upstream optimal trajectory is planned regardless of the risk of running a red light. However, this risk is eliminated by introducing a risk assessment procedure, where a critical stopping distance $d_{cr}(t)$ is calculated at every time $t$. The vehicle policy is set to only decelerate at the maximum allowable deceleration level $\alpha$ if the remaining distance to the stop bar is less than or equal to $d_{cr}(t)$ (3). Which is a similar approach to what was used in the literature [17].

$$X_S = \sum_{t=t_0}^{\mathbb{E}(t_s)} v(t, x(t)) \cdot \Delta t \leq X_M \quad (2)$$

$$v(t + \Delta t, x(t + \Delta t)) = v(t, x(t)) + \alpha \cdot \Delta t \quad \forall \{(x,t): t < \mathbb{E}(t_s), X_M - x \leq d_{cr}(t)\} \quad (3)$$

Similarly, the downstream speed policy is constrained to cover the downstream distance plus the remaining distance that was not covered by the vehicle in the case where the signal switches before the vehicle reaches the position $X_M$ (4). Note that this constraint is applied after the true switching time is revealed.

$$\sum_{t=t_s}^{t_f} v(t, x(t)) \cdot \Delta t = X_N + (X_M - X_S) \quad (4)$$

The speed policy is further constrained by the kinematic equation (5), where $a(t)$ is the vehicle acceleration / deceleration level. Equation (6) shows the speed limit constraint. Finally, the acceleration policy is constrained by the maximum jerk limitation of $1.3\frac{m}{s^2}$ to ensure the passengers' comfort (7). [18]

$$v(t + \Delta t) = v(t) + a(t) \cdot \Delta t \quad \forall t \in [t_0, \ldots, t_f] \quad (5)$$

$$v(t) \leq v_{lim} \forall t \in [t_0, \ldots, t_f] \quad (6)$$

$$a(t + \Delta t) \leq a(t) + 1.3 \cdot \Delta t \quad \forall t \in [t_0, \ldots, t_f], \forall \text{ throttle level} > 0 \quad (7)$$

## V. UNDERLYING SYSTEMS

Vehicle dynamics models are utilized to define the spatiotemporal variables according to the current state variables and the acting forces on the vehicle including the tractive, aerodynamic, rolling, and grade resistance forces. As the solution space is discretized in time and space, vehicle dynamics and fuel consumption models are used to evaluate the optimal vehicle trajectory at each time step $\Delta t$.

### A. Vehicle Dynamics Model

Based on the throttle level or braking inputs, it is required to model the vehicle's acceleration and deceleration resulting behavior considering all the acting forces and constraints on the vehicle. As the scope of this research is optimizing the trajectory of light-duty vehicles, the dynamic model for light-duty vehicles on varied terrain was used [19]. The vehicle acceleration, tractive, and resistance forces are computed in this model as follows:

- The vehicle acceleration is computed by dividing the net force acting on the vehicle, which is the tractive force minus the resisting force, by the vehicle mass $m$ (8). Where $F(t)$ and $R(t)$ are the tractive and resistance forces at time t, respectively.

$$a(t) = (F(t) - R(t))/m \quad (8)$$

- The vehicle tractive force is computed as the acting force by the engine. This force is upper bounded by the maximum tractive force between the vehicle tires and the roadway pavement (9). Where $f_b$ is the throttle input (from 0 to 1), $\eta_d$ is the driveline efficiency, β is the factor that accounts for gear shift impacts (set to 1.0 for light-duty vehicles) [19]. $P(t)$ is the vehicle power at time $t$, $v(t)$ is the vehicle speed at time $t$, $M_{ta}$ is the vehicle mass on the tractive axle (kg), μ is the road friction or adhesion coefficient. and $g$ is the gravitational acceleration in (m/s²).

$$F(t) = \min\left[3600 f_b \eta_d \beta \frac{P(t)}{v(t)}, M_{ta} g \mu\right] \quad (9)$$

- The acting resistance force on the vehicle is the summation of the vehicle rolling, aerodynamic, and grade resistance forces (10). Where $\rho$ is the air density, $C_d$ and $C_h$ are the vehicle drag coefficient and the altitude correction factor, respectively. $A_f$ is the vehicle frontal area. $c_{r0}, c_{r1}$ and $c_{r2}$ are the rolling resistance constants.

$$R(t) = \frac{\rho}{25.91} C_d C_h A_f v^2(t) + mg\frac{c_{r0}}{1000}(c_{r1}v(t) + c_{r2}) + mgG(t) \quad (10)$$

### B. Fuel Consumption Model

It is required to estimate accurate rates of fuel consumption levels consistent with actual in-field measurements. To achieve that, the Virginia Tech Comprehensive Power-Based Fuel Consumption model (VT-CPFM-1) is used [20]. This model overcomes the limitations of existing fuel consumption models in literature by eliminating the bang-bang control behavior. In addition, it is easily calibrated using publicly available vehicle data. It is also known for its simplicity and accuracy in calculating instantaneous fuel consumption from the instantaneous vehicle power. Further details about the model can be found in the literature [20]. The formulation of the used model is provided below (11):

$$FC(t) = \begin{cases} \alpha_0 + \alpha_1 P(t) + \alpha_0 P(t)^2 & \forall P(t) \geq 0 \\ \alpha_0 & \forall P(t) < 0 \end{cases} \quad (11)$$

Where $\alpha_0, \alpha_1$ and $\alpha_2$ are the model constant calibrated for the specific vehicle in use. $P(t)$ is the instantaneous vehicle power calculated (12).

$$P(t) = \left(\frac{R(t) + 1.04\, ma(t)}{3600\eta_d}\right) v(t) \quad (12)$$

Here $a(t)$ and $v(t)$ are the instantaneous acceleration and speed variables, respectively, $m$ is the vehicle mass, and $\eta_d$ is the driveline efficiency. As the vehicle used in the field test is a 2014 Cadillac SRX, we used this vehicle's calibrated parameters so that we can compare the simulated results with the field experiment results as a baseline for our calculations. The vehicle parameters are provided in TABLE II.

TABLE II
CADILLAC SRX 2014 PARAMETERS

| Parameter | Value | Parameter | Value |
|---|---|---|---|
| $\alpha_0$ | 7.89E-04 | $A_f$ | 3.33 |
| $\alpha_1$ | -5.77E-19 | $\eta_d$ | 0.92 |
| $\alpha_2$ | 2.27E-06 | % Mass on tractive axle | 0.54 |
| $c_{r0}$ | 1.75 | $m$ | 2388 |
| $c_{r1}$ | 0.0328 | $P_{max}$ | 229.7 |
| $c_{r2}$ | 4.55 | $C_h$ | 0.95 |
| $c_d$ | 0.39 | $\rho$ | 1.2256 |

V. SOLUTION APPROACH

*A. Stochastic Dynamic Programming*

Given the non-linear stochastic optimization setting in this particular problem, finding the optimal speed policy is a computationally complex procedure. This complexity needs to be reduced to enhance the real-world applicability of this algorithm. So that we can find, in a feasible time frame, a heuristic solution that is close enough to the optimal one. Dynamic programming (DP) is one of the most powerful methods to solve stochastic optimization problems in discretized time utilizing Bellman's principle of optimality, and it is well known for its significant reduction of computation complexity [21].

In our problem, we have a single control variable ($m = 1$), which is either a throttle input or a deceleration level. The problem is solved in a control time horizon of length $T = (t_f - t_0)/\Delta t$ (where $t_0$ is the simulation starting time, and the $t_f$ is the final time when the vehicle reaches the destination at the position $X_N$). Thus, the solution space would be $\mathbb{R}^{m \times T}$. DP will decompose the problem to a sequence of $T$ problems on $\mathbb{R}^m$, which is a significant reduction in computational complexity.

As depicted in Fig. 3, the solution approach is described as follows:
1. The vehicle enters the upstream link with an initial state at position $S_0$ while the signal is red.
2. Based on the current speed, the system calculates the critical distance $d_{cr}$ required to decelerate at the maximum allowable deceleration level ($\alpha = -6\ m/s^2$). In addition, the system checks the remaining distance to the signalized intersection stop line $D_i$.
3. While the signal is red, the system evaluates the risk of running the red light. The risk is evaluated by comparing the remaining distance to the intersection with the critical stopping distance. In this case, there are two possibilities:
   a. The red-light violation risk occurs if the remaining distance is less than or equal to the critical distance. In this case, a deceleration policy with the rate $\alpha$ is adopted.
   b. If there is no violation risk, the system will receive SPaT information, generate the next state policy for the vehicle using the A* algorithm, and repeat steps 1 through 4.
4. If the signal turns green, the system will generate the next downstream policy states until reaching the destination at the position $S_M$ using the A* algorithm.

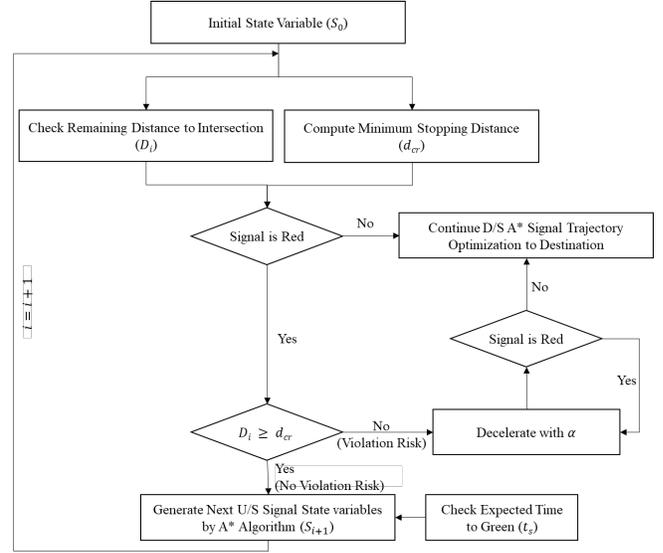

**Fig. 3.** The Stochastic Setting Solution System

*B. The A* Algorithm*

The A* algorithm is a pathfinding algorithm that uses a heuristic cost estimate to determine the next state policy leading to the shortest path. It was used previously in the literature to plan the vehicle trajectory in deterministic settings [12]. The A* algorithm is used to estimate the least-cost next-state acceleration/deceleration policy by assuming this policy would remain the same for the remainder of the time horizon (Fig. 4).

In our problem, the cost estimate heuristic is calculated over two sections; the upstream and the downstream sections described as follows:
1. In the upstream section, the system iterates in an outer loop for each next-state upstream admissible policy, the algorithm assumes the policy would remain the same until the reaching the position $X_s$. The policies that violate the expected red-light condition are considered infeasible. The system computes the upstream section fuel consumption $U_i$ for each feasible policy $i$.
2. For each upstream feasible policy $i$, the system iterates in an inner loop generating the next downstream admissible policies based on the vehicle state at position $X_s$. Similarly, each admissible downstream policy is assumed to remain the same until reaching the destination at the position $X_N$. The system computes the downstream section fuel consumption $D_{ij}$ for each upstream policy $i$ and downstream policy $j$.
3. Using the upstream and downstream heuristic cost estimates, the A* algorithm selects the upstream next state

policy based on the minimum total fuel consumption of the two sections $U_i + D_{ij}$ for each upstream policy $i$.

4. When the vehicle reaches the downstream section, the system will only iterate in the inner loop, and the policy with minimum fuel consumption $D_{ij}$ is selected.

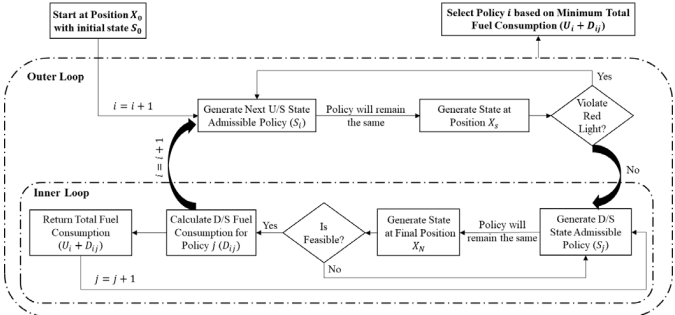

**Fig. 4.** Logic of A-Star Algorithm to Find the Minimum Path

## VI. RESULTS & ANALYSIS

### A. Optimal Policy for Deterministic SPaT information

In the case of deterministic SPaT information, the trajectory optimization system is applied for our problem in the defined scenarios described in TABLE I. As the system receives the signal switching time from the controller, the optimal trajectory is computed to achieve the best fuel economy. As shown in Fig. 5 and Fig. 6, at TTG=10 seconds, the optimal policy is to cruise at the current speed. That is because the travel time to the intersection at the current speed would be greater than the time to green. As such, the system applies the minimum throttle level that overcomes the resistance forces that keep the vehicle cruising at the current speed and keeps the fuel losses at a minimum.

For the case of TTG=15 seconds, the system recognizes that some delay is required to reach the stop line after the signal switches. Accordingly, the system applies a level of deceleration to achieve that delay. Similarly, at TTGs of 20 and 25 seconds, the system applies the optimal deceleration levels to achieve the required delay. Downstream the traffic signal, the system determines the optimal acceleration policy to reach the destination position with the desired speed for all cases.

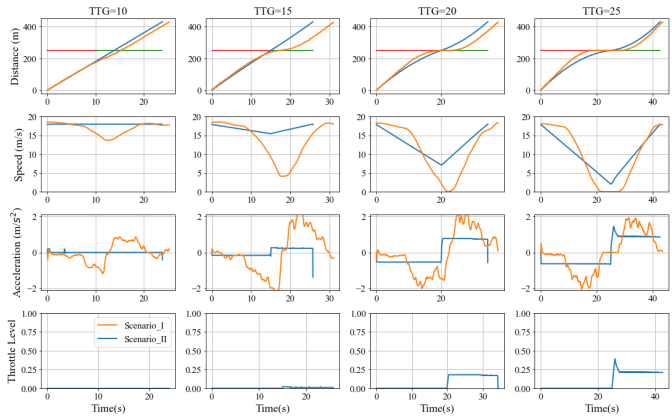

**Fig. 5.** Example trajectory plots for downhill grade showing the difference in traveled distance, speed, and acceleration profiles between the uninformed driver (Scenario I) and the optimal deterministic trajectory (Scenario II). The optimal throttle level profile is also shown.

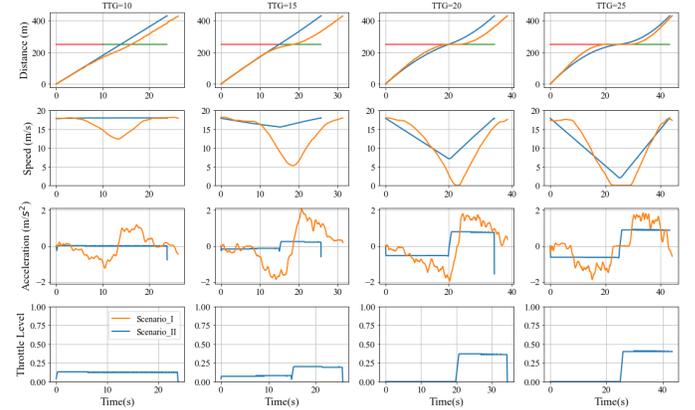

**Fig. 6.** Example trajectory plots for uphill grade showing the difference in traveled distance, speed, and acceleration profiles between the uninformed driver (Scenario I) and the optimal deterministic trajectory (Scenario II). The optimal throttle level profile is also shown.

### B. Optimal Policy for Stochastic SPaT information

In the case the exact signal switching time is not provided, signal switching time predictions are used to plan the vehicle trajectory. In our simulations, these predictions are generated to mimic the stochastic SPaT data received from an actual traffic signal controller. This information is assumed to follow a constrained normal distribution, where the prediction uncertainty is represented by two components: a bias value added to the mean of the distribution, and a standard deviation that reflects the random errors in the SPaT prediction model (SD). The bias and SD values are sampled from zero to a maximum predefined value of 8 seconds. At each $\Delta t$, a new random distribution is generated based on the vehicle position.

The system is applied for the defined scenarios in TABLE I with different distributions, where a sampled value from the probability distribution of the switching time is transmitted to the system at each $\Delta t$. The system generates the policy that achieves the best fuel economy according to the perceived information at this point, which generates a fluctuating policy behavior as shown in Fig. 7 for the downhill setting and Fig. 8 for the uphill setting. The system controls the fluctuation intensity based on the jerk limit, such that the maximum allowable jerk limit is not violated to ensure proper passenger comfort levels.

In the case at TTG=10 seconds, the system's optimal policy is to cruise with the current initial speed until reaching the risk region, where the vehicle decelerates to stop as long as the signal is still red. This risk assessment procedure is represented by the sudden drop in the acceleration profile (Fig. 7). When the traffic signal switches to green, the vehicle accelerates again to reach the destination position at the desired speed.

Similarly, for the case at TTG=15 and 20 seconds, the system generates the optimal deceleration policy, while continuously checking the risk region. If the risk region is reached, the vehicle will activate the deceleration policy to stop as long as the signal is still red. If the signal switches to green during deceleration, the vehicle will update its policy to accelerate again to reach the desired speed at the destination position.

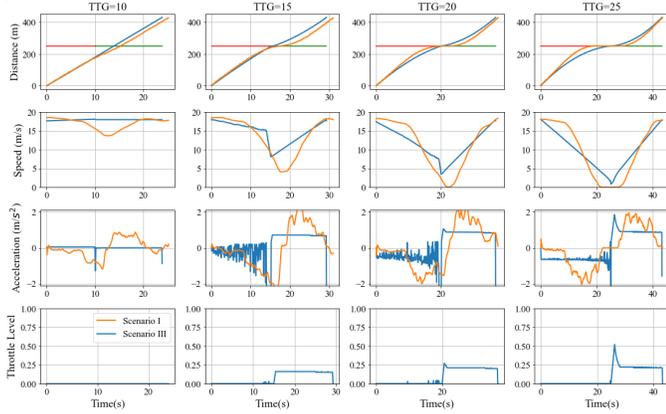

**Fig. 7.** Example trajectory plots for downhill grade showing the difference in traveled distance, speed, and acceleration profiles between the uninformed driver (Scenario I) and an optimal stochastic trajectory (Scenario III). The optimal throttle level profile is also shown.

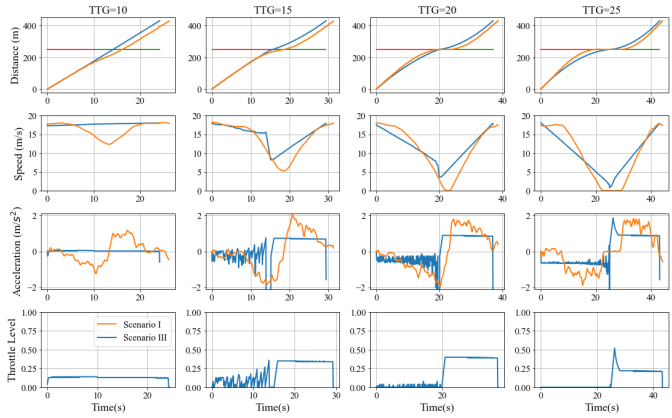

**Fig. 8.** Example trajectory plots for uphill grade showing the difference in traveled distance, speed, and acceleration profiles between the uninformed driver (Scenario I) and the optimal stochastic trajectory (Scenario III). The optimal throttle level profile is also shown.

*C. Fuel Consumption Savings for Deterministic and Stochastic Settings*

Compared to the uninformed driver, the optimization system showed significant savings in fuel consumption levels for both the deterministic and stochastic settings. The results shown compare the overall performance in terms of fuel savings in the case when the SPaT information is deterministic (Scenario II) to when SPaT information is stochastic (Scenario III) using the uninformed driver (Scenario I) as a baseline. The results for the stochastic SPaT are obtained by averaging the results for all the different levels of bias and standard deviations used. As depicted in Fig. 9, the overall average fuel savings are 37% and 28% for the deterministic and stochastic settings, respectively. The fuel savings can reach up to 63% for deterministic TTG and 42% for stochastic TTG in the scenario where the car is moving downhill with TTG=15 seconds. The reason why fuel savings are maximized with downhill grade is that minimal throttle is needed to overcome the resisting forces and make the vehicle cruise. The reason why TTG= 15 is the case where most savings are realized compared to uninformed drivers is that this is the speed where a very small deceleration is required to reach the stop bar by the time it turns green. This is because the average speed needed to reach the stop bar at the time it turns green is 37.3 mph (16.67 m/s) and the starting speed is 40 mph (17.88 m/s). Accordingly, the optimal policy would be to slow down a little bit to reach the stop bar at a high speed. In a stochastic setting, the policy becomes more conservative with more than needed deceleration to account for the possibility that the SPaT information is not 100 percent correct. An uninformed driver will not have this information and will be more likely to stop. This demonstrates that the closer the time to green is to the time to reach the stop bar given the vehicle speed, the more sensitive the fuel consumption is to the accuracy of the prediction.

The case of TTG = 25 sec is when the switching time is too long for the vehicle to arrive at the stop bar without stopping, it is implied that the saving in fuel consumption is exclusively due to optimizing the acceleration profile in the downstream section. That is demonstrated by the fact that the fuel consumption savings for deterministic and stochastic SPaT information over the uninformed driver are almost the same (21% in scenario II and 20% in scenario III). As the switching time decreases, a notable difference between the two scenarios is achieved, as optimizing the trajectory in the upstream section plays an effective role in decreasing the total fuel consumption. It is shown that optimizing the upstream trajectory incorporates significant portions of additional fuel savings that can be achieved. These savings amount to up to 22% for deterministic SPaT and 19% for stochastic SPaT, which demonstrates the importance of having reliable predictions for traffic signal switching times.

*D. Effect of Bias and Variance*

As the switching time prediction quality impacts the optimal trajectory, a sensitivity analysis is conducted to visualize the impact of normal distribution parameters (the error bias and standard deviation) on the optimal fuel consumption levels. A total of 320 runs were performed using different values of bias and standard deviation (SD) (Fig. 10). The analysis showed that the change in SD values has a significant effect on fuel consumption, compared to the effect of the bias value, which was not shown to be significant. Additional uncertainty in the prediction led to losses in fuel savings except in the case where TTG=15s.

Fig. 11 shows that in the case of TTG=15s, the fuel consumption decreases as the stochasticity increases. This behavior is explained by considering that the vehicle's travel time from the initial position to the intersection equals the switching time, according to the initial speed of 40 mph. The vehicle approaches the stop line at a time when the signal is about to switch to green. However, the system activates the risk assessment procedure to ensure that no violations of traffic signal timings are made. Therefore, the vehicle begins to decelerate until the signal switches to green at some point. By introducing additional stochasticity, the vehicle can be delayed enough so that the signal will switch before the vehicle enters the risk zone and activates the stopping strategy. That is the only case where the more stochastic the information is, the better fuel economy would be.

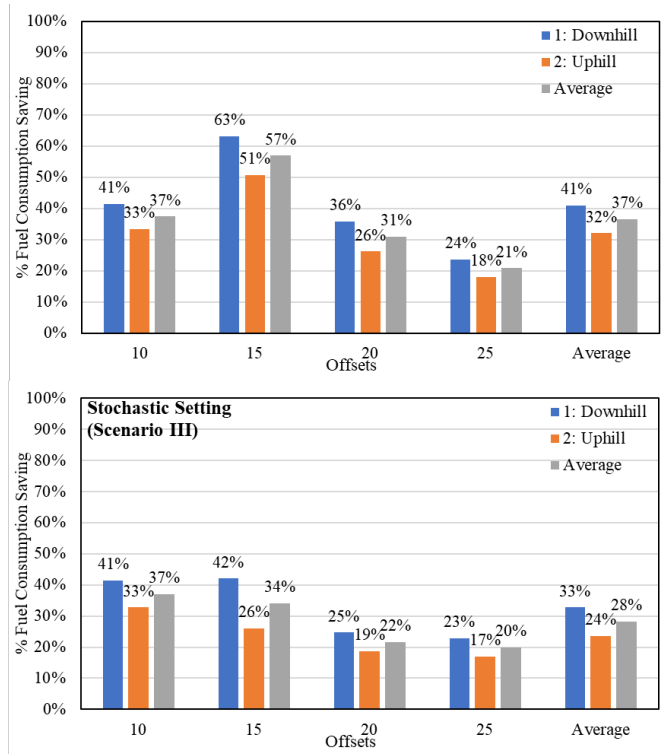

**Fig. 9.** Fuel Consumption Savings of Scenario II and III over Scenario I

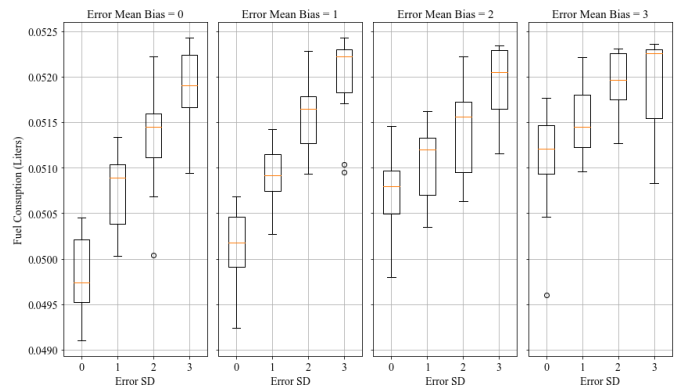

**Fig. 10.** Fuel Consumption Values versus Different Bias and Standard Deviation (SD) Values in the case of Time to Green (TTG) offset equals 20 seconds, Downhill Vertical Grade.

*E. The Effect of Varying Initial Speed*

To further evaluate the system response to the stochastic SPaT information, we vary the initial vehicle speed from 40 mph to 29 mph (Fig. 12). For the slower initial speed, the system generates almost a constant policy regardless of the switching time prediction uncertainty in the cases where TTG is 10 and 15 seconds. The policy is to provide a minimal level of constant acceleration such that the vehicle reaches the stop bar within the green and reaches the destination at the maximum possible speed (40 mph). That is because the optimal policy given the lower speed will have the vehicle arrive at the intersection late enough to accommodate the switching time probability distribution in the cases where TTG is 10 and 15 seconds, respectively. In cases where TTG is 20 and 25 seconds, the system will generate a regularly fluctuating acceleration policy to decrease the speed of the vehicle enough to reach the intersection by the time the signal turns green. Note that the slow deceleration is followed by a sharp deceleration as the vehicle is within the safe stopping distance from the stop bar. This highlights the previously mentioned point that the prediction accuracy becomes more important the closer the time to green is to the time until reaching the intersection given the initial speed.

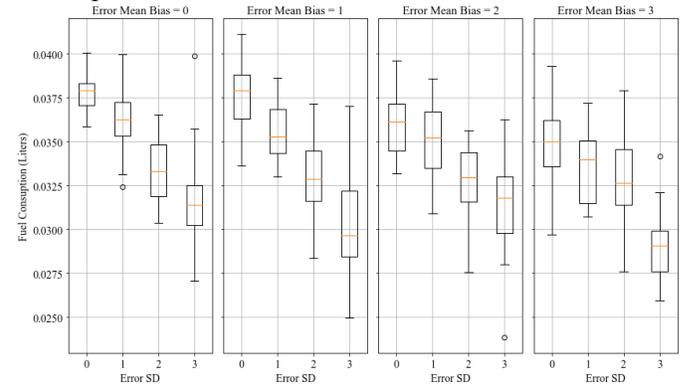

**Fig. 11.** Fuel Consumption Values versus Different Bias and Standard Deviation (SD) Values in the case of Time to Green (TTG) offset equals 15 seconds, Downhill Vertical Grade.

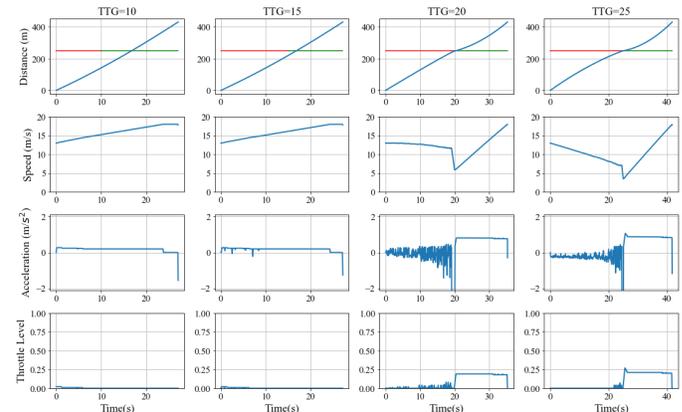

**Fig. 12.** Optimal trajectory plots of the vehicle's initial speed equal 13m/s (29 mph)

*F. Impact of Confidence in SPAT Information on the Fuel Consumption*

This section discusses the level of confidence in the switching time prediction required to achieve significant fuel consumption savings in the stochastic setting of the optimal velocity planning problem. Given that the fuel consumption saving in the deterministic setting is the upper bound of the stochastic setting, we need to identify how the level of uncertainty decreases the saving. In other words, how the probability distribution standard deviation and bias can impact the fuel savings compared to the upper bound. Error! Reference source not found. shows the relationship between the parameters of the switching time probability distribution and the proportion of fuel saving, where the maximum saving occurs in the case when there is no uncertainty in the switching time information (prediction SD and mean bias equal zero). The saving proportion decreases as these parameters increase. It is shown that we can achieve fuel savings in the stochastic setting of more than 85% of the maximum possible savings when the switching time distribution standard deviation is less than 1.25 seconds, and the mean bias is less than 0.8 seconds. In other words, 85% of the possible fuel savings can be achieved with the switching time prediction error up to ±3.3 seconds for a 95% confidence level. This information gives us insights into the required confidence in SPaT prediction to achieve significant fuel savings using the optimal speed planning system proposed in this paper.

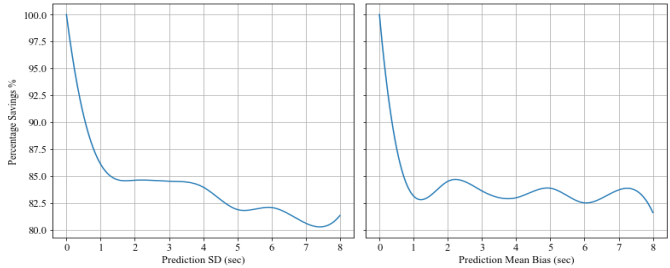

**Fig. 13.** Effect of Standard Deviation and Bias on Fuel Consumption Savings Compared to the Maximum Possible Savings Achieved in Scenario II

## V. CONCLUSION & RECOMMENDATIONS FOR FURTHER RESEARCH

An optimal Green Light Optimal Speed Advisory (GLOSA) system is developed to find the optimal trajectory for the vehicle approaching a traffic signal controller. The system deals with the fixed time as well as the actuated traffic signals, where the exact signal switching time is unknown and a probability distribution for the switching time is provided to the vehicle through the I2V. Using fuel consumption minimization as the objective function, the problem is solved through a Dynamic Programming (DP) procedure utilizing the A-Star algorithm to find the minimum-cost path. A risk assessment procedure is implemented to control the vehicle acceleration and deceleration levels so that the red light is not violated. Additionally, the passengers' comfort is achieved by controlling the vehicle acceleration jerk to limit the disruption due to the error fluctuations in the expected switching time at each time step.

The simulation results of deterministic and stochastic switching time settings are compared with the case of the uninformed driver experiment conducted at Virginia Tech Transportation Institute. Results show that significant fuel savings can be achieved with an average of 37% and 28% for the deterministic and the stochastic settings, respectively. Additionally, the sensitivity analysis showed that the system is robust to the errors included in the uncertain switching time predictions, where the system will adjust the vehicle trajectory in real time according to the updated predicted probability distribution of the switching time. It also showed how initial speed affects the vehicle acceleration and deceleration fluctuations which can be useful in planning trajectories in a corridor that has consecutive signal controllers.

In the case of uncertain switching times, the proposed system can achieve more than 85% of the possible savings achieved in the case of fixed time signals if the timing error is (± 3.3 seconds) at a 95% confidence level. It was shown that the system is more sensitive to prediction errors when the time to green is close to the time required for the vehicle to reach the intersection given its current speed.

As is the case with any study, further research is needed. Future directions include testing the stochastic optimizer for different levels of market penetration within surrounding uninformed drivers when queues form upstream of the traffic signal, and for different levels of traffic congestion, as was done in the case of deterministic signal timings [22-24].

ACKNOWLEDGMENT

This effort was funded by the University Mobility and Equity Center (UMEC).

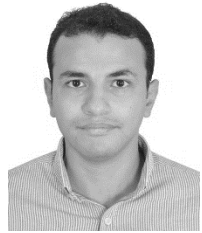

**Amr K. Shafik** received the B.Sc. degree (Hons.) in civil engineering from the Cairo University, Cairo, Egypt, in 2016, and the M.Sc. degree in Transportation Engineering from Cairo University in 2020. He is currently pursuing the Ph.D. degree with Charles E. Via, Jr. Department of Civil and Environmental Engineering. He works as a Graduate Research Assistant with the Center for Sustainable Mobility, Virginia Tech Transportation Institute. His research interests include traffic control and optimization, traffic modeling and simulation, intelligent transportation systems, and transportation planning.

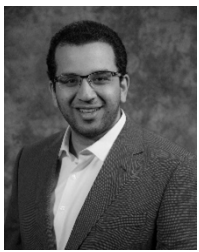

**Seifeldeen Eteifa** received his B.Sc. degree (Hons.) in construction engineering and management from the American University in Cairo, Cairo, Egypt, in 2016, and the M.Sc. degree in Construction Engineering and Technology from the University of Tennessee, Knoxville in 2018. He is currently pursuing the Ph.D. degree with Charles E. Via, Jr. Department of Civil and Environmental Engineering. He works as a Graduate Research Assistant with the Center for Sustainable Mobility, Virginia Tech Transportation Institute. His research interests cover applications of machine learning in intelligent transportation systems, modeling and simulation, construction and transportation safety and social network analysis.

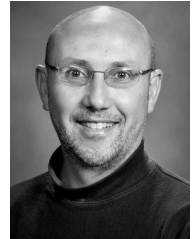

**Hesham A. Rakha** (M'04, SM'18, F'20) received the Ph.D. degree from Queen's University, Kingston, Ontario, in 1993. He is currently the Samuel Reynolds Pritchard Professor of Engineering in the Department of Civil and Environmental Engineering and the Department of Electrical and Computer Engineering (Courtesy) at Virginia Tech, and the Director of the Center for Sustainable Mobility at the Virginia Tech Transportation Institute. His research focuses on large-scale transportation system optimization, modeling, and assessment. He works on optimizing transportation system operations, including vehicle routing, developing various network and traffic signal control algorithms, developing freeway control strategies (speed harmonization and ramp metering), and optimizing vehicle motion (lateral and longitudinal control of connected automated vehicles (CAVs)) to enhance their efficiency and reduce their energy consumption while ensuring their safety. Dr. Rakha is a Fellow of the IEEE. He is the recipient of IEEE ITS Outstanding Research Award (2021) from the IEEE Intelligent Transportation Systems Society. He was the author/co-author of six conference best paper awards, namely: 19th ITS World Congress (2012), 20th ITS World Congress (2013), VEHITS (2016), VEHITS (2018), and TRB (2020); received the most cited paper award from the *International Journal of Transportation Science and Technology (IJTST)* in 2018; and received 1st place in the IEEE ITSC 2020 UAS4T Competition. In addition, Dr. Rakha received Virginia Tech's Dean's Award for Outstanding New Professor (2002), the College of Engineering Faculty Fellow Award (2004-2006), and the Dean's Award for Excellence in Research (2007). He is an Editor for *Sensors* (the Intelligent Sensors Section), an Academic Editor for the *Journal of Advanced Transportation*, a Senior Editor for the *IEEE Transactions of Intelligent Transportation Systems*, and an Associate Editor for the *Journal of Intelligent Transportation Systems: Technology, Planning and Operations* and the *SAE International Journal of Sustainable Transportation, Energy, Environment, & Policy*. Furthermore, he is on the Editorial Board of the *Transportation Letters: The International Journal of Transportation Research* and the *International Journal of Transportation Science and Technology*.